\documentclass[sigconf,screen]{acmart}
\usepackage[utf8]{inputenc}

\usepackage{amsmath}
\usepackage{bm}
\AtBeginDocument{%
  \providecommand\BibTeX{{%
    Bib\TeX}}}

\usepackage{amsmath,amssymb,amsfonts}
\usepackage{algorithmic}
\usepackage{multicol}
\usepackage{graphicx}
\usepackage{textcomp}
\usepackage{xcolor}
\usepackage{tabularx}
\usepackage{tcolorbox}
\usepackage{colortbl}
\usepackage{longtable}
\usepackage{lipsum}
\def\BibTeX{{\rm B\kern-.05em{\sc i\kern-.025em b}\kern-.08em
    T\kern-.1667em\lower.7ex\hbox{E}\kern-.125emX}}
\usepackage{multirow}
\usepackage{enumitem}
\setlist[itemize,1]{left=0pt}

\setcopyright{acmcopyright}
\copyrightyear{2025}
\acmYear{2025}
\acmDOI{}
\acmISBN{}
\acmConference {SCORED Workshop of CCS}{October 17, 2025}{Taipei, Taiwan}
\acmBooktitle{SCORED Workshop of ACM CCS Conference, October 17, 2025, Taipei, Taiwan}



\copyrightyear{2025}
\acmYear{2025}
\setcopyright{acmlicensed}\acmConference[SCORED '25]{Proceedings of the 2025 Workshop on Software Supply Chain Offensive Research and Ecosystem Defenses}{October 13--17, 2025}{Taipei, Taiwan}
\acmBooktitle{Proceedings of the 2025 Workshop on Software Supply Chain Offensive Research and Ecosystem Defenses (SCORED '25), October 13--17, 2025, Taipei, Taiwan}
\acmDOI{10.1145/3733827.3765526}
\acmISBN{979-8-4007-1915-8/2025/10}

\begin{document}

\title{Establishing a Baseline of Software Supply Chain Security Task Adoption by Software Organizations}

\author{Laurie Williams}
\affiliation{
  \institution{North Carolina State University }
  \city{ Raleigh}
  \state{ North Carolina}
  \country{USA }
}
\email{ lawilli3@ncsu.edu}

\author{Sammy Migues }
\affiliation{
  \institution{Imbricate Security }
  \city{ Sterling}
  \state{VA }
  \country{USA }
}
\email{sammy.migues@gmail.com }

\begin{abstract}

Software supply chain attacks have increased exponentially since 2020. The primary attack vectors for supply chain attacks are through: (1) software components; (2) the build infrastructure; and (3) humans (a.k.a software practitioners). Software supply chain risk management frameworks provide a list of tasks that an organization can adopt to reduce software supply chain risk.  Exhaustively adopting all the tasks of these frameworks is infeasible, necessitating the prioritized adoption of tasks.  Software organizations can benefit from being guided in this prioritization by learning what tasks other teams have adopted.  The goal of this study is to aid software development organizations in understanding the adoption of security tasks that reduce software supply chain risk through an interview study of software practitioners engaged in software supply chain risk management efforts. An interview study was conducted with 61 practitioners at nine software development organizations that have focused efforts on reducing software supply chain risk.  The results of the interviews indicate that organizations had implemented the most adopted software tasks before the focus on software supply chain security. Therefore, their implementation in organizations is more mature. The tasks that mitigate the novel attack vectors through software components and the build infrastructure are in the early stages of adoption.  Adoption of these tasks should be prioritized.    
\end{abstract}

\maketitle

\section{Introduction}
Software organizations did not anticipate how, or how quickly, the software supply chain (SSC) would become a deliberate attack vector. 
Attackers have moved from finding and exploiting vulnerabilities contributed by well-intentioned developers, such as log4j~\cite{log4J,csrb2022Log4j}, to actively implanting malicious vulnerabilities directly into software dependencies~\cite{Roadmap}. 
Recent high-profile supply chain attacks, such as SolarWinds~\cite{cisa2021Joint,solarwindsadvisory,mandiantsolarwinds} and XZ Utils~\cite{XZOpenWallFreund,XZRedHatSecurityAdvisory,xztimeline}, made headlines and directed attention towards the importance of SSC security and to software security, in general. 
Sonatype reported that between November 2023 and November 2024, 512,847 malicious packages were discovered in open-source software ecosystems, a 156\% increase year-over-year growth~\cite {sonatype_2024}.  

In response to this growth in software supply chain attacks and increasing cyber-attacks and resulting harms, worldwide government actions have caused a flurry of initiatives by software organizations to: implement good software security practices within their software development lifecycle; to impose good software security practices on their vendors; and to establish measurement frameworks that enable the attestation of the use of these practices. Most prominently, Section 4 of US Executive Order (EO) 14028~\cite{EO_2021}, issued May 12, 2021, calls for enhancing the security of software products purchased by the US government through the adoption by software developers of software security practices. 
The White House Office of Management and Budget (OMB) subsequently issued the M-22-18 memo~\cite{EO_memo} (\textit{Enhancing the Security of the Software Supply Chain through Secure Software Development Practices}) documenting requirements for software organizations that supply software to the US government to provide self-attestation of the use of secure software development practices.  
Similarly, the European Union (EU) released the Cyber Resilience Act (CRA)~\cite{cra}, which introduces mandatory cybersecurity requirements for software manufacturers, including software design and development requirements.  
The implications of not complying with the CRA include significant fines, the inability to sell a software-based product in the EU, and reputational damage.

Software supply chain security frameworks (e.g.,~\cite{souppaya2022secure, NIST_800161, BSIMM, scvs, Microsoft_framework, SLSA, CNCF_TAG, Scorecard}) list tasks that an organization can adopt to reduce software supply chain risk.  Some of the frameworks also provide guidance on planning a secure software supply chain risk management initiative. Note, we use the term ``task'', also synonymous with security control, practices, safeguards, and countermeasures~\cite{NISTSecurityControls}. Security tasks are actions, procedures, techniques, or other measures that an organization can adopt to reduce the vulnerability of an information system~\cite{NISTSecurityControls}. Examples of software supply chain security tasks for code originating from outside the organization include determining whether the project is actively maintained, conducting code reviews before merging pull requests, and utilizing static, dynamic, and software composition analysis (SCA) tools as detection measures for vulnerabilities and malicious code.

While there is some overlap in these frameworks, they still collectively constitute a lengthy list of tasks.  As a result, organizations must prioritize the adoption of SSC risk management (SSCRM) tasks because every organization has time and resource limitations, as well as limitations on the amount of change the organization \textemdash ~especially the product security team and the software engineering teams \textemdash~can handle at one time. As a result, organizations can benefit from guidance for doing this prioritization, making them keenly interested in learning about what other organizations are doing, particularly companies like themselves, when adopting a new technology or business approach~\cite{Crossing, Redwine}.

As such, \textit{the goal of this study is to aid software development organizations in understanding the adoption of security tasks that reduce software supply chain risk through an interview study of software practitioners engaged in SSCRM efforts.}  To achieve our goal, we first pose the following research question:

\vspace{-1.8mm}
\begin{tcolorbox}[colback=magenta!5!white,colframe=magenta!75!black]
    \textbf{RQ:  What software supply chain security tasks are adopted in security-focused companies?}
\end{tcolorbox}
\vspace{-2mm}



An interview study was conducted with 61 practitioners in software development groups at nine organizations that have focused efforts on improving software supply chain security and/or complying with EO 14028, such that they can sell software-based products to the US government. The task list and interview protocol used were based upon the assessment questions of the Proactive Software Supply Chain Risk Management (P-SSCRM)~\cite{williams2024proactive} framework because it comprehensively unifies 10 prominent software supply chain frameworks. After the completion of each interview, tasks were scored based on the progress of adoption as discussed in the interviews.  We summarize our observations of efforts within those nine companies.  

This paper contributes the following:
\begin{itemize}
    \item Empirical results of the adoption of tasks to reduce software supply chain risk by nine security-focused software companies.  These results establish a baseline that can be built upon/compared with studies of other companies and/or additional longitudinal studies with the same companies.
    \item A reusable interview framework to enable other researchers to perform similar studies of task adoption, centered around the assessment questions from the P-SSCRM.
    \item An invitation to other researchers to collaborate on studies of task adoption, such that a family of research results can be established that can aid in the prioritization of task adoption by companies. 
\end{itemize}

Section~\ref{RelatedWork} of this paper provides information on related work.  Section~\ref{ResearchMethodology} provides our research methodology.  We present the results of the research in Section~\ref{Results} and the takeaways of the study in Section~\ref{Takeways}. We conclude with the limitations of the work in Section~\ref{Limitations}, the ethical considerations in Section~\ref{Ethics}, and the summary and future work in Section~\ref{summary}.

\section{Related Work} \label{RelatedWork}
This section summarizes related work for this research.

\subsection {Building Security In Maturity Model} \label{BSIMM}
The design of our study is based on the seventeen years of success of the Building Security In Maturity Model (BSIMM) \cite{BSIMM}.  
In 2008, Gary McGraw, Sammy Migues, and Brian Chess created the BSIMM as a \textit{descriptive} model to allow external assessment of organizations' state-of-practice in secure software development. The BSIMM framework is built around the assessment of the activities an organization may adopt in support of its software security initiative. Each assessment evaluates which of the now 128 software security ‘activities’
have been adopted by the organization. 

Annual BSIMM reports provide high-level findings to the industry based on descriptive analysis. The latest such report is freely available~\cite{BSIMM}. Each report provides detailed descriptions of the activities; a grouping of the activities into 12 practices; and adoption percentages for each activity.  Generally, we use the BSIMM assessment and reporting methodology.  However, BSIMM assesses organizations as adopting or not adopting security activities (a binary value).  As will be discussed in Section \ref{interviews}, we use five values to indicate where an organization is in its adoption journey to a finer level of granularity than BSIMM. 

\subsection {Software Security Practice Frameworks} \label{frameworks}

Several frameworks, in addition to BSIMM, provide guidance or enumeration of software security practices.  
The OWASP Software Assurance Maturity Model (SAMM)~\cite{OWASP_samm} is an open prescriptive model to help organizations assess, formulate, and implement, through a self-assessment model, a strategy for software security practice adoption that is tailored to the specific risks facing the organization.   

The US National Institute of Standards and Technology (NIST) Cybersecurity Framework (CSF)~\cite{CSF} provides organizations with a structure to aid in understanding and improving cybersecurity risk. 
The US Department of Defense (DoD) Cybersecurity Maturity Model Certification (CMMC)~\cite{CMMC} framework combines cybersecurity standards and best practices and maps 171 practices into five maturity levels that range from basic cyber hygiene to advanced optimization. 

The ISO/IEC 27001~\cite{ISOIEC27001} and NIST SP 800-53~\cite{NIST_80053} standards provide requirements for establishing, implementing, operating, monitoring, maintaining, and improving the security of a digital information management system.  
The ISO/IEC 27034 Application Security standard~\cite{ISO27034} offers definitions, concepts, principles, and non-prescriptive guidance to help organizations integrate security into the processes for managing their applications. The OWASP Application Security Verification Standard (ASVS)~\cite{ASVS} provides developers with a list of requirements for secure development.

The above frameworks are in common use.  We mention them in this section to highlight that many software security frameworks have minimal mentions of software supply chain security risk.  As such, they broadly inform SSCRM thinking but are not included in the P-SSCRM model.   

\subsection {Task and mapping artifacts} \label{Mapping}

Government and industrial organizations have developed frameworks ~\cite{souppaya2022secure, NIST_800161, BSIMM,scvs, Microsoft_framework,SLSA, CNCF_TAG, Scorecard,EO_2021, Attestation} that enumerate tasks to reduce software supply chain risk by all the roles in an organization involved in reducing that risk.  These frameworks each have a different focus.  For our study, we were looking for a comprehensive set of tasks that can reduce software supply chain risk.  As will be discussed in Section~\ref{P-SSCRM}, the P-SSCRM framework is the union of the tasks from these frameworks and provides a mapping between all the frameworks when a task appears in more than one framework.

The Open Source Security Foundation (OpenSSF) of the Linux Foundation has created the Open Source Project Security Baseline (OSPS Baseline) to provide controls to help project maintainers understand security best practices and expectations~\cite{OSSFBaseline}. Consumers of OSPS Baseline can assess their project and use the assessment results to understand how their usage of the tasks impacts their own security and compliance goals. OSPS Baseline maps its tasks to seven external frameworks, including the OpenSSF Best Practices Badge (BPB)~\cite{BPB}, the CSF~\cite{CSF}, the Cyber Resilience Act (CRA)~\cite{cra}, the Software Security Development Framework (SSDF)~\cite{souppaya2022secure}, ISO/IEC 18974 (OC)~\cite{OC}, the Open Cybersecurity Reference Architecture (OpenCRE)~\cite{ocre}, the Supply Chain Levels for Software Artifacts (SLSA)~\cite{SLSA}, and the P-SSCRM~\cite{williams2024proactive}. OpenCRE also links together and maps various security standards and guidelines.   However, OSPS Baseline and OpenCRE are broader (i.e., software security) and are not focused on software supply chain security.  Additionally, our study focuses on both proprietary and open-source software and organizations, whereas the OSPS Baseline focuses solely on open-source software.  Additionally, the original version of P-SSCRM (v1.0) and predates the release of OSPS Baseline.    

Lee et al.~\cite{lee2025softwaresecuritymappingframework} introduce the Software Security Mapping Framework to operationalize security requirements.  The framework systematically maps 131 requirements to 400 actionable operational steps spanning the software development lifecycle. This framework is also broader than is needed for our study on reducing risk in software supply chain security.  


\section{Research Methodology } \label{ResearchMethodology}

In this section, we describe our research methodology.  In Section \ref{P-SSCRM}, we provide background on the P-SSCRM framework we used to structure our interview study. We discuss our interview process in Section \ref{interviews} and our data analysis process in Section \ref{DataAnalysis}.   

\subsection{Proactive Software Supply Chain Risk Management Framework (P-SSCRM)} \label{P-SSCRM}

 The Proactive-Software Supply Chain Risk Management Framework (P-SSCRM)~\cite{williams2024proactive} provides the union and restructuring of the tasks outlined in ten frameworks~\cite{souppaya2022secure, NIST_800161, BSIMM,scvs, Microsoft_framework,SLSA, CNCF_TAG, Scorecard,EO_2021, Attestation}.  These ten frameworks were selected due to their emphasis on software supply chain security and because each contributed to the development of a holistic view of software supply chain risks by software teams.  Securing the software supply chain is a team effort of practitioners in Governance, Business Management, Software Security, Development, Architecture, DevOps, and IT. P-SSCRM contains tasks for all of these roles.  
 
 P-SSCRM is organized as follows:
 \begin{itemize}
     \item \textbf{Task}:  Actions or efforts conducted for the goal of implementing a secure software application and of reducing the security risk of that application and for its producing organization.  Each task has a lower-level objective to aid in secure software development and risk reduction. P-SSCRM consists of \textbf{73 tasks}.  For example:  P.3.1: Component and container choice:  make informed third-party component and container choices.  For each of the 73 tasks, P-SSCRM provides (1) a unique identifier; (2) a unique name; (3) the objective of the task; (4) a description of the task; (5) assessment questions; and (6) the mapping to the framework(s) that contributed to the task.
     \item \textbf{Practice}: A grouping of P-SSCRM tasks that have similar mid-level objectives to aid in secure software development and risk reduction. The 73 tasks are organized into \textbf{15 practices}.  For example:  P.3 Manage component and container choices:  software supply chain risk can be reduced by careful choice and handling of third-party components and containers.
     \item \textbf{Group}:  A grouping of P-SSCRM practices with a similar high-level objective to aid in secure software development and risk reduction.  The 15 practices of the P-SSCRM are organized into \textbf{four groups:  Governance (G), Product (P), Environment (E), and Deployment (D)}.  For example:  Product (P):  Tasks that lead to deploying a secure product with minimal vulnerabilities, with associated required attestations and artifacts.
 \end{itemize}

  \begin{table}[tb]
\caption{Task coverage by contributing framework [brackets indicate the framework contributes a unique task]}\label{TAB:TaskCoverage}
\begin{tabular}{l l l l l l}
\toprule
Framework & Gov & Prod & Env & Deploy & \#/Total   

\\ \midrule

P-SSCRM             & 23 & 19 & 22 & 8 & 73/73                 \\ 
EO/SSDF             & 11 & 14 & 4 & 5 & 34/34                          \\ 
Self-Attest             & 8 & 12 & 4 & 5 & 23/34 (SSDF)                      \\ 
BSIMM             & 17 [1] & 14  & 2 & 4 & 37/125\footnote{BSIMM contains software security activities that do not involve software supply chain.}\\ 
SLSA             & 2 & 1 & 3 & 0 & 6/6              \\ 
800-161             & 20 [5] & 10 & 9 & 5 [1] & 44/183\footnote{44 of the 183 tasks are specifically identified in NIST 800-161r1 as mapping to Executive Order 14028 and included in P-SSCRM. The scope of NIST 800-161r1 includes manufacturing/hardware. }                        \\ 
SCVS             & 1 & 5 & 5 & 0 & 11/11                     \\ 
S2C2F             & 3 & 7 [1] & 3 & 2 [1] & 15/15             \\ 
CNCF            & 4 & 6 & 13 [8] & 1 [1] & 24/24         \\ 
Scorecard            & 0 & 6 & 2 & 1 & 9/9\footnote{OpenSSF Scorecard has 18 metrics.  Fifteen (15) of these mapped to 9 tasks, with five activities having more than one OpenSSF metric.  }         \\ 
 
\bottomrule
\end{tabular}
\end{table}

 Business managers, including risk managers and vendor managers, and a software security group generally conduct the tasks in the Governance group.  Architects, developers, and testers conduct the tasks in the Product group. Information Technology (IT) and DevOps teams conduct tasks within the Environment groups.  DevOps and the Product Security Incident Response Team (PSIRT) perform the tasks in the Deployment group.     

 As each of the ten standards~\cite{souppaya2022secure, NIST_800161, BSIMM,scvs, Microsoft_framework,SLSA, CNCF_TAG, Scorecard,EO_2021, Attestation} was considered for inclusion in the P-SSCRM, the strengths of each and their value in working together synergistically to provide a holistic view of software supply chain risk reduction by all five roles were considered. The number of tasks in each of the four groups that came from each framework is shown in Table \ref{TAB:TaskCoverage}.
 
 Twenty of the 23 Governance tasks came from the NIST 800-161 framework. The Business Manager and Software Security roles often conduct the practices of the governance group, for which NIST 800-161 provides broad coverage of P-SSCRM tasks in this group. Fourteen of the 19 Product tasks came from the SSDF framework. Product tasks are done by the Architect/Developer role, and the SSDF is a development framework. Thirteen of the 23 Environment tasks came from the CNCF SSC framework. With its focus on the cloud environment, CNCF SSC framework contains practices to protect the build infrastructure and computing environment.  Finally, 5 of 8 tasks from the Deployment group come from SSDF. The purpose of including the OpenSSF Scorecard metrics in the table is to provide a longitudinal status of the ability of software ecosystems to automatically detect evidence that a task has been conducted on a software product/project.  Numbers in square brackets indicate the number of tasks that appeared only in the contributing framework. 

Similar to BSIMM~\cite{BSIMM}, the P-SSCRM framework provides the structure for a ``descriptive'' model.  That is, P-SSCRM is not a prescriptive model that specifies what an organization should do to reduce software supply chain risk. Rather, using the structure of P-SSCRM, we can provide information on what the organizations that have undergone a P-SSCRM assessment are doing.  Similarly, P-SSCRM is descriptive, as we expect the data to become more important over time. In contrast, a prescriptive model would require annual updates and would only address the common denominator.

\subsection{Interviews}\label{interviews}
The P-SSCRM framework ~\cite{williams2024proactive} includes assessment questions for each of the 73 tasks that were used to conduct the interviews.  These assessment questions were utilized to conduct an empirical study of software supply chain security task adoption among nine industry-leading software supply chain risk management initiatives in two medium-sized companies (hundreds of employees) and seven large companies (thousands of employees) in the United States and Europe.  The identities of the companies remain anonymous due to non-disclosure agreements (NDA) associated with the interviews\footnote{The NDA precludes the release of the interview records and prevents another reviewer from analyzing the interview notes post hoc.}. These nine organizations have mature Software Security groups and focused efforts on reducing software supply chain risk, and were interested in an assessment of their software supply chain risk management initiative.   Hence, the industry-leading organizations selected could serve as a guide to other organizations with less mature initiatives. 

The first author conducted semi-structured interviews with 5 to 13 practitioners in each of the nine companies\footnote{The interviews were conducted in 2022-2023.}.
Each interview lasted an average of 75 minutes~to assess the adoption of the subset of the 73 tasks based on the job role of the interviewee using the assessment questions in the P-SSCRM framework.  
In most cases, more than one practitioner at a company was asked about each task, enabling cross-checking of answers within one organization. The interviews varied from the P-SSCRM assessment questions to probe inconsistent answers within and between interviewees.    

A total of 61 practitioners were interviewed:  1 Chief Information Security Office (CISO), 27 in the general area of Governance (software security group, risk management, vendor management); 23 in the Product area (architect, developer); and 10 in the areas of Environment/Deployment (DevOps, PSIRT).  Table \ref{TAB:Roles} provides a breakout of the job roles of the interviewees for the nine companies.  Note that the count of interviewees for the Environment and Deployment roles was combined because these interviewees could answer questions in both areas.Obtaining input from a variety of roles provided different perspectives and expertise in the interviews, enriching the assessment conducted. The one CISO and 13 of the 27 interviewees in the Governance roles were from the software security group.  These 14 interviewees often had a broad perspective of task adoption across the organization. 

\begin{table}[tb]
\caption{Job roles of interviewees for each company}\label{TAB:Roles}
\begin{tabular}{l l l l l l}
\toprule
Company & CISO & Gov & Prod & Env \& Deploy & Total   

\\ \midrule

1             & 0 & 2 & 4 & 0 & 6                 \\ 
2             & 0 & 4 & 1 & 1 & 6                          \\ 
3             & 0 & 4 & 1 & 2 & 7                      \\ 
4             & 0 & 3  & 3 & 0 & 6\\ 
5             & 0 & 3 & 1 & 2 & 6              \\ 
6             & 0 & 2 & 2 & 1 & 5                        \\ 
7             & 1 & 2 & 0 & 1 & 4                     \\ 
8             & 0 & 4 & 8 & 1 & 14             \\ 
9            & 0 & 3 & 3 & 1 & 7         \\ 
Total            & 1 & 27 & 23 & 10 & 61         \\ 

\bottomrule
\end{tabular}
\end{table}

The interviews were transcribed.  Upon completion of all the interviews for a company, each task was assigned a value based on the company's stage in the adoption journey.  Each task was assigned one of five values. These values and the typical indications/expressions from the interviews for each are as follows:   
\begin{itemize}
    \item \textbf{0:  Not observed}: "We haven't started doing that (yet)," "We plan to work on that next year," "That task is not a priority for us."
    \item \textbf{0.25:  Emerging}: "We have one team piloting that task," "We are starting to look into what it would take to do that."
    \item \textbf{0.50:  Progress being made}: "We purchased a tool, and several engineers have been using it."  "We are getting the kinks out of our use of that task."
    \item \textbf{0.75:  Close to implementing}: "That task is fully adopted by two teams, and we're spreading it to more teams now."  
    \item \textbf{1.00:  Task implemented}:  "We've been doing that for years."  "We have focused on implementing that task over this past year, and all teams are doing that now."
\end{itemize}  

\begin{figure}[t]
  \centering
      \includegraphics[width=0.48\textwidth, height=1.8\textheight, keepaspectratio]{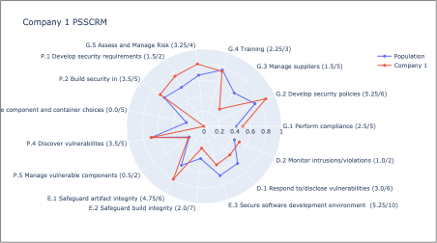}
  \caption{
    \textbf{Sample spiral diagram for a company }
  }
  \label{fig:Spiral}
\end{figure}

Each company was provided a report summarizing their interviews.  The report included: (1) the numerical value (0, 0.25. 0.5. 0.75, 1.00) for each of the 73 tasks; (2) a comparison of their task values to the average value for all nine companies; (3) a qualitative assessment of strengths and weakness of the organization; and (4) graphics summarizing the organizations average practice adoption versus the full set of nine companies. A sample spiral diagram is provided in Figure \ref{fig:Spiral}. The values assigned to each task (similar to Table \ref{tab:AllTasks}) were reviewed with representatives of the companies.  Through these conversations, the value assigned to each task converged between the interviewer and the company representative whereby the representative was able to provide the required descriptive evidence to support a change in value or the interviewee shared comments from the interviews to support the value.  Company representatives were most interested in the gap analysis of the value of the tasks and practices for their company versus all nine companies, such as found in Figure \ref{fig:Spiral}.

\subsection{Data Analysis}\label{DataAnalysis}
We used the adoption data from the 61 interviews on 73 tasks to answer our research question:  \textbf{RQ:  What software supply chain security tasks are adopted in security-focused companies?} We leverage descriptive quantitative statistics to analyze the data with associated qualitative observations.
For each company, the adoption values for each task in a practice were averaged to produce a practice average for each of the 15 practices.   Additionally, an average for each of the 15 practices was computed as the average across the nine companies.

\section {Results} \label{Results}

  In this section, we present the summary results from the interviews.  In Section \ref{task}, we present average results for the 73 tasks and 15 practices.  In Sections \ref{top} and \ref{bottom}, we discuss the most and least adopted tasks, respectively.

\subsection{Adoption by Task and Practice}\label{task}
 
Table~\ref{tab:AllTasks} provides the average adoption rate for all nine companies for each of the 73 tasks. Each row contains the P-SSCRM identifier and an abbreviated name for the task.  In the identifier, the Group of practices can be found by the first letter (G=Governance, P=Practice, E=Environment, D=Deployment). For each of the 73 tasks, an organization received a score between 0 and 1, inclusive.  They get a 0 if the organization has not started adopting the task, a 0.25 if their adoption is emerging, a 0.5 if progress is being made, a 0.75 if the practice is almost adopted, and a 1.0 if the practice is adopted. 

\begin{figure*}[t]
  \centering
  \includegraphics[width=.8\linewidth, keepaspectratio]{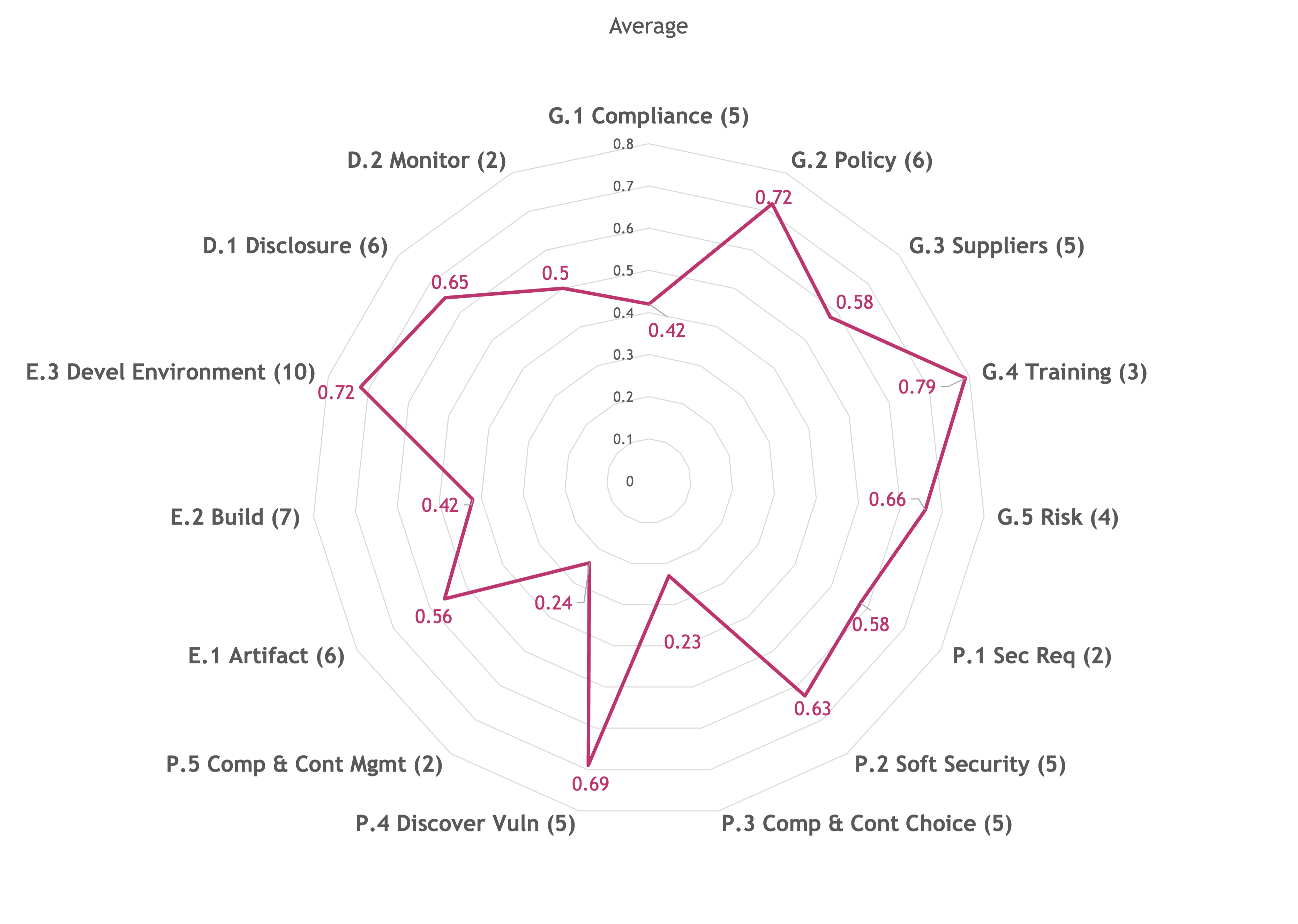}
  \caption{Adoption by practice for nine companies}
  \label{fig:AllSpiral}
\end{figure*}

The polar chart in Figure~\ref{fig:AllSpiral} summarizes the adoption of each of the 15 practices by all 9 companies by providing the average of the tasks within the practice.   If the average is toward the center of the circle, the organizations have collectively not adopted the tasks in the practice. If the average is toward the outside of the chart, the organizations are doing pretty well with adoption.

Some practices in Figure~\ref{fig:AllSpiral} have lower adoption.  The two lowest are the P.3 Component and Container Choice with an average of 0.23 for the five tasks in this practice and P.5 Component and Container Management with an average of 0.24 for the two tasks in this practice. With averages of 0.23 and 0.24, the tasks in these practices were beginning to emerge in implementation at the 9 companies. Without the implementation of the tasks in P.3 and P.5, the attack vector through the components in the supply chain is likely still available to attackers. As shown in Figure~\ref{fig:AllSpiral}, the third lowest practice is E.2 Safeguard Build Integrity with an average 0.42 (indicating that some progress is being made toward adoption) for the seven tasks of this practice.  Without the adoption of these tasks, the attack vector through the build infrastructure is likely still available to attackers. Six of the bottom 11 tasks adopted are from these three practices, as will be discussed in Section \ref{bottom}.   

Alternatively, the practice of G.4 Training has the highest adoption with an average of 0.79 for these three tasks. The six tasks in the G.2 Security Policy and the E.3 Secure Development Environment practice have higher adoption, with an average of 0.72 for both of these practices. Interviews indicate that organizations have been working on task adoption in these areas before the focus on software supply chain security. Five tasks in these three practices appear in the top ten adopted tasks, as discussed in Section \ref{top}.   

\begin{table*}[t]
    \centering
    \caption{Average task adoption by all companies for 73 P-SSCRM tasks. The full definition for each task can be found at \cite{williams2024proactive}}
    \label{tab:AllTasks}
    \begin{tabular}{|lr|lr|lr|lr|}
    \toprule
    \multicolumn{2}{|c|}{\cellcolor{gray!30!white} \textbf{Governance (G)}} & \multicolumn{2}{c|}{\textbf{\cellcolor{gray!30!white} Product (P)}} & \multicolumn{2}{c|}{\textbf{ \cellcolor{gray!30!white} Environment (E)}} & \multicolumn{2}{c|}{\textbf{\cellcolor{gray!30!white} Deployment (D)}}\\
    \cellcolor{gray!30!white} \textbf{Task} & \cellcolor{gray!30!white} \textbf{Avg} & \cellcolor{gray!30!white} \textbf{Task} & \cellcolor{gray!30!white} \textbf{Avg} & \cellcolor{gray!30!white} \textbf{Task} & \cellcolor{gray!30!white} \textbf{Avg} & \cellcolor{gray!30!white} \textbf{Task} & \cellcolor{gray!30!white} \textbf{Avg} \\
    \cmidrule(lr){1-2} \cmidrule(lr){3-4} \cmidrule(lr){5-6} \cmidrule(lr){7-8}
     \textbf{G.1 Compliance} & & \textbf{P.1 Secure Requirements} & & \textbf{E.1 Artifact Integ}& & \textbf{D.1 Vuln Response} & \\
    G.1.1 Org requiremts & 0.75 & P.1.1 Prod requiremts & 0.61 & E.1.1 Release artifacts& 0.72 & D.1.1 Vuln analysis & 0.78 \\
    G.1.2 Sftw license & 0.89 & P.1.2 Release integrity& 0.56 & E.1.2 Version control& 0.81 & D.1.2 Vuln remediation& 0.83 \\
    G.1.3 Prod attestation& 0.17 &   & & E.1.3 MFA& 0.53 & D.1.3 Vuln disclosure& 0.86 \\
    G.1.4 Deliv provenance& 0.08 & & & E.1.4 Developer SSH& 0.44 & D.1.4 Vuln eradicate & 0.44 \\
    G.1.5 Deliver SBOM & 0.19 &  &  & E.1.5 Branch protection& 0.53 & D.1.5 Emerg artifact& 0.28 \\
    
    & & & & E.1.6 Decomm assets& 0.47 & D.1.6 Root cause anal& 0.69 \\
  &&&&&&&\\
   \textbf{G.2 Secure Policy} & & \textbf{P.2 Build Security In} & & \textbf{E.2 Build Integrity}& & \textbf{D.2 Monitor Intrusn} & \\ 
    G.2.1 Upper mgmt & 0.83 & P.2.1 Design review & 0.58 & E.2.1 Release policy& 0.33 & D.2.1 System monitor& 0.81  \\
    G.2.2 Secure SDLC& 0.72 & P.2.2 Secure coding& 0.69 & E.2.2 Verify depend& 0.28&D.2.2 Build monitoring& 0.18  \\
    G.2.3 Roles \& Resp& 0.92  & P.2.3 Secure-by-default& 0.50 & E.2.3 Defensive compil& 0.25& & \\
    G.2.4 Security features& 0.58 & P.2.4 Vetted repositories& 0.69 & E.2.4 CI/CD Hosting& 0.47 & & \\
    G.2.5 Asset inventory& 0.42 & P.2.5 In-house compont& 0.69  & E.2.5 Secure orchestr& 0.64 & & \\
    G.2.6 Info at rest & 0.86 & & & E.2.6 Reproduc builds& 0.03 & & \\
    & & & & E.2.7 Build output& 0.94 & & \\
    &&&&&&&\\
     \textbf{G.3 Manage Suppliers} & & \textbf{P.3 Comp \& Cont Choice} & &E.3 \textbf{Secure Dev Env} & & & \\
    G.3.1 Security contract& 0.58  & P.3.1 C\&C choice& 0.39 & E.3.1 Authentication& 1.00 & & \\
    G.3.2 Sep of duties& 0.64 & P.3.2 Trust repos& 0.25& E.3.2 Env separation& 0.84 & & \\
    G.3.3 Info disclosure& 0.64 & P.3.3 Signed commits & 0.08 & E.3.3 RBAC& 0.81 & & \\
    G.3.4 Session audits& 0.50 & P.3.4 Vetted repositories& 0.31& E.3.4 Info flow enforce& 0.67 & & \\
    G.3.5 Notification agree& 0.56 & P.3.5 Vetting bypass& 0.14  & E.3.5 Baseline config& 0.58 & & \\
     & & &  & E.3.6 Config settings& 0.81 & & \\
    & & & & E.3.7 Boundary protect & 0.91 & & \\
    &  & & & E.3.8 Key rotation & 0.69 & & \\
       & & & & E.3.9 Ephemeral cred & 0.22 & & \\
     & & & & E.3.10 Root trust & 0.66 & & \\
       &&&&&&&\\
     \textbf{G.4 Training} & & \textbf{P.4 Discover Vuln} & & & & & \\
    G.4.1 Role training& 0.97& P.4.1 Security code rev& 0.58 & &  & & \\
    G.4.2 Conting training& 0.61& P.4.2 Scanning tools& 0.97 & &  & & \\
    G.4.3 Attack trends & 0.72& P.4.3 Automated detect & 0.64 & &  & & \\
    & & P.4.4 Security testing& 0.69 & &  & & \\
    & & P.4.5 3rd party compl& 0.58 & &  & & \\
      &&&&&&&\\
     \textbf{G.5 Risk Assessment} & & \textbf{P.5 Comp \& Container Manag} & & & & & \\
    G.5.1 Criticality analy & 0.58 & P.5.1 SBOM consump & 0.03&  &  & & \\
    G.5.2 Security risks& 0.69 & P.5.2 Depend update & 0.44& &  & & \\
     G.5.3 Security metrics & 0.72 & & & & & & \\
    G.5.4 Data-inform dec& 0.67 & & & & & & \\
    \bottomrule
    \end{tabular}
\end{table*}


\begin{table}[tb]
\caption{Top ten tasks adopted}\label{TAB:PSSCRM_Adoption}
\begin{tabular}{l l l}
\toprule
\textbf{ID} & \textbf{Task Name} & \textbf{Average Adoption}                                    \\ \midrule
E.3.1             & Authentication & 1.00                 \\ 
P.4.2             & Automated security scanning tools & 0.97                          \\ 
G.4.1             & Role-based training & 0.97                      \\ 
E.2.7             & Build output & 0.94  \\ 
G.2.3             & Roles and responsibilities & 0.92              \\ 
E.3.7             & Boundary protection & 0.91                         \\ 
G.1.2             & Software license & 0.89                     \\ 
G.2.6             & Protection of information at rest & 0.86               \\ 
D.1.3             & Vulnerability disclosure & 0.86         \\ 
E.3.2             & Environmental separation & 0.84      \\ 
\bottomrule
\end{tabular}
\end{table}

\begin{table}[!htb]
\caption{Bottom eleven tasks adopted}\label{TAB:PSSCRM_Least_Adoption}
\begin{tabular}{l l l}
\toprule
\textbf{ID} & \textbf{Task Name} & \textbf{Average Adoption}                                    \\ \midrule
E.2.6             & Reproducible builds & 0.03                 \\ 
P.5.1             & SBOM Consumption & 0.03                          \\ 
P.3.3            & Require signed commits & 0.08                      \\ 
G.1.4             & Deliver provenance & 0.08  \\ 
P.3.5            & Prevent component vetting bypass & 0.14              \\ 
G.1.3             & Produce attestation & 0.17                         \\ 
D.2.2             & Build process monitoring & 0.18                     \\ 
G.1.5             & Deliver SBOM & 0.19               \\ 
E.3.9             & Ephemeral credentials & 0.22         \\ 
E.2.3  &   Defensive compilation and build & 0.25     \\ 
P.3.2  &   Trusted repositories & 0.25     \\ 
\bottomrule
\end{tabular}
\end{table}

\subsection{Top 10 Tasks Adopted}\label{top}
Table~\ref{TAB:PSSCRM_Adoption} provides the average adoption for the Top 10 most adopted tasks. Four of the Top 10 are in the Environment group, indicating a maturity in securing development and build environments by the IT and DevOps roles.  Empirical observations suggest that IT and DevOps organizations had already begun hardening their development and build environments before the current focus on software supply chain security.    

The highly adopted Governance tasks are (1) conducting role-based security training (G.4.1) and (2) establishing roles and responsibilities for handling security within an organization (G.2.3) conducted by the Software Security Group; and (3) identifying software licenses that may conflict with an organization's policies conducted by the Business Manager role (G.1.2); and (4) protection of information at rest (G.2.6).  Sammak et al. \cite{DomYas} conducted 18 semi-structured interviews with experienced developers.  Their results indicate a lack of defined security roles and insufficient security training.  These inconsistencies may be explained by the companies in the P-SSCRM study having focused efforts on reducing software supply chain security risk.  Software organizations increase their legal risks if they utilize software that conflicts with their organization's policies.  Therefore, SCA tools have historically been used to check license compliance.  SCA tool functionality has expanded in recent years to identify vulnerabilities in these software components and to produce a Software Bill of Materials (SBOM).  As such, SCA tools are playing a more significant role with the focus on software supply chain security.  

The only Product task in the Top 10 is the use of automated security scanning tools (P.4.2), which many organizations are integrating into their CI/CD pipelines and are using to scan third-party components.  Finally, the only Deployment task in the Top 10 is vulnerability disclosure (D.1.3).  Organizations are enhancing their vulnerability disclosure programs and are understanding the implications of the bad press that comes when a vulnerability is not disclosed.  In general, the Top 10 Tasks were adopted by organizations before the focus on software supply chain security, and therefore, their implementation in organizations is more mature.

\subsection{Bottom 11 Tasks Adopted}\label{bottom}

Table~\ref{TAB:PSSCRM_Least_Adoption} provides the average adoption rate for the Bottom 11 least adopted tasks.  Eleven (instead of 10) are chosen due to a tie between the bottom two tasks.  In the US, the Governance tasks of producing and delivering attestation (G.1.3), provenance (G.1.4), and SBOM (G.1.5) are becoming mandatory, though adoption of these tasks is low.  Consuming these artifacts is not yet happening, as indicated by the interviews and the SBOM consumption task (P.5.1). Among interviewees, these tasks may still be considered compliance ''checkboxes'' that are not yet strictly enforced. Xia et al.   \cite{sbom} conducted a mixed qualitative and quantitative study for gathering
data from 17 interviewees and 65 survey respondents from 15
countries across five continents to gain an understanding of how practitioners perceive SBOMs and the
challenges of adopting SBOMs.  Their study also found a lack of SBOM consumption and highlighted a need for better tooling to enable SBOM consumption.  

The Prevent component vetting bypass (P.3.5) task is in the Bottom 11, indicating that even a limited approval process for components and containers may still be easily bypassed.  Also, the use of trusted repositories, such as npm or DockerHub (P.3.2), is low.  Often, developers are free to upload a component or container from any source.   

Three of the Environment tasks appear in the Bottom 11.  The reproducible build task (E.2.6) has not been fully implemented by any company.  This bit-by-bit comparison between two or more builds of the same product is conducted to confirm that no malicious backdoor injections have taken place during the build process.  Organizations can see the value in reproducible builds, but the time and resources needed to build multiple times have kept this task from being implemented.  Companies are also not using compiler, interpreter, and build tool features to reduce vulnerabilities, such as producing compiler warnings for vulnerable code that are treated as errors or applying obfuscation techniques (E.2.3).   Finally, organizations have also not adopted the task of enhancing the security of the development environment by using ephemeral credentials (E.3.9).

The Deployment task contained in the Bottom 11 list is Build Process Monitoring (D.2.2) in which an organization continuously monitors the build and deployment environment to detect suspicious or unexpected activity, such as attempts to connect to suspicious endpoints during the build process. Build process monitoring could enable the discovery of tampering in the build process, as was done with the SolarWinds attack.

\subsection{Contextualizing the Results with Other Studies}\label{Nusrat}

Organizations may base their task adoption strategy on a variety of information. This paper examines the interest an organization has in the tasks adopted by similar organizations.  Other studies have presented alternative approaches. In this section, we contextualize our results with these studies. 

\subsubsection{Related to Security Outcomes}
Organizations may want to prioritize the adoption of tasks based on empirical evidence showing that adopting tasks improves security outcomes and mitigates security risks.  Multiple large-scale empirical studies analyzing open-source repositories ~\cite{zahan2023openssf, zahan2023software, zahan2025prioritizing}~investigated the relationship between software security tasks and security outcome metrics. These studies, involving data analysis of the npm ecosystem, found that the OpenSSF Scorecard Code-Review, Maintained status, Pinned Dependencies, and Branch Protection are the tasks most strongly associated with security outcomes. Table \ref{TAB:Scorecard_Adoption} provides the adoption of these four tasks using the mappings provided in the P-SSCRM framework. With adoption percentages in the 0.5 range, three of these tasks are in the "Progress being made" range.  The task embodied by the Maintained metric is higher, indicating that organizations are considering whether a component/container is actively maintained in their choices.

\begin{table}[tb]
\caption{Adoption of tasks associated with security outcomes}\label{TAB:Scorecard_Adoption}
\begin{tabular}{l l l}
\toprule
\textbf{Scorecard Metric} & \textbf{P-SSCRM Task} & \textbf{Avg Adoption \%}                                    \\ \midrule
Code Review             & P.4.1 & 0.58                 \\ 
Maintained             & P.3.1 & 0.83                          \\ 
Pinned dependencies             & G.2.4 & 0.58                      \\ 
Branch protection             & E.1.5 & 0.53  \\ 
 
\bottomrule
\end{tabular}
\end{table}

\subsubsection{Mitigating Current Attacker Techniques}
Organizations may wish to prioritize the adoption of tasks that mitigate some specific attack techniques that attackers are using.  Hamer et al. ~\cite{hamer2025closing}, systematically mapped the attack techniques in three prominent software supply chain attacks (SolarWinds ~\cite{cisa2021Joint,solarwindsadvisory,mandiantsolarwinds}, Log4j~\cite{log4J,csrb2022Log4j}, and XZ Utils~\cite{XZOpenWallFreund,XZRedHatSecurityAdvisory,xztimeline} to mitigating framework tasks. The research presented a "Starter Kit" of the most important 10 tasks to adopt to mitigate these attack techniques. Table \ref{TAB:Sivana} provides the adoption progress for these ten tasks. As can be seen, six of these ten have an average adoption of 0.81 or higher, indicating the nine organizations have made significant progress in adopting these tasks that mitigate a current attacker trend. Organizations should prioritize the adoption of P.5.2 (Dependency update) and P.2.1 (Security design review) since the average adoption of these tasks is lower.  Note:  E.3.11 (Environmental scanning tools) was not included in P-SSCRM v1.01 because it was not included in any of the ten contributing frameworks.  Hamer et al. are working with framework authors to include this task in future versions of the contributing frameworks.  

Two tasks are contained in both the Top 10 Most Adopted task list and the Top 10 Mitigating Task list.  Organizations may want to prioritize the adoption of these tasks since most other organizations have adopted these tasks, and these tasks have been shown to mitigate the attack techniques used in the Solarwinds, Log4j, and XZ Utils attacks.  These tasks are G.2.6 (Protection of information at rest) and E.3.7 (Boundary protection).  

\begin{table}[tb]
\caption{Adoption of ranked tasks associated with mitigating current attacker techniques ~\cite{hamer2025closing}}\label{TAB:Sivana}
\begin{tabular}{l l }
\toprule
\textbf{Task} & \textbf{Avg Adoption}                                    \\ \midrule
E.3.3  Role-based access control             & 0.81                 \\ 
D.2.1  System monitoring            & 0.81                          \\ 
E.3.7  Boundary protection            & 0.91                      \\ 
E.3.6  Monitor changes to config. settings            & 0.81   \\ 
E.3.11 Environmental scanning tools              &  New                 \\ 
P.5.2  Dependency update             & 0.44                          \\ 
P.2.1  Security design review             & 0.58                       \\ 
E.3.4  Information flow enforcement            & 0.67   \\ 
G.2.6  Protection of information at rest             & 0.86                       \\ 
D.1.2  Risk-based vulnerability remediation            & 0.82   \\ 
\bottomrule
\end{tabular}
\end{table}
\subsubsection{BSIMM15 Results}  In 2025, the BSIMM15 \cite{BSIMM} report was published.  As discussed in Section \ref{BSIMM}, annual BSIMM reports provide high-level findings to the industry based on descriptive analysis of companies that have been assessed. The BSIMM15 report summarizes the results of the adoption of 128 software security 'activities' by 121 assessed organizations (e.g. BSIMM15 CR1.4 was adopted by 88\% of the 121 organizations).  Recall from Table \ref{TAB:TaskCoverage}, 37 of these 128 activities are mapped to 73 P-SSCRM tasks.  Over the past 17 years, BSIMM has provided the only set of comprehensive results of the adoption of software security activities in the software industry.    

In Tables \ref{TAB:BSIMM_Top 10Adoption},  \ref{TAB:BSIMM_Least_Adoption}, and \ref{TAB:BSIMM:  Sivana}, we compare the P-SSCRM adoption results for the most and least adopted P-SSCRM tasks and the top 10 mitigating tasks, respectively. The purpose of the comparison is twofold.  First, we aimed to compare the BSIMM15 adoption results with those in the P-SSCRM study to determine if the results converge. Second, we aim to determine if the noteworthy P-SSCRM tasks (top 10, bottom 11, mitigating) are already covered in the well-established BSIMM. This analysis will provide feedback on whether P-SSCRM assessments are necessary to generate research results on software supply chain risk management, or if the BSIMM results are sufficient.

\begin{table}[tb]
\caption{BSIMM:  P-SSCRM-ranked (highest to lower) top ten tasks adopted}\label{TAB:BSIMM_Top 10Adoption}
\begin{tabular}{l l l}
\toprule
\textbf{P-SSCRM ID} & \textbf{BSIMM15 Mapping} & \textbf{BSIMM15 Adoption \%}                                    \\ \midrule
E.3.1             & None & N/A                 \\ 
P.4.2             & CR1.4, ST1.4, ST2.5 & 88, 45, 51                          \\ 
G.4.1             & T1.1, T1.7, T1.8, T2.5, & 51, 50, 41, 34\\ & T2.8, T2.9, T3.1, T3.2 & 21, 26, 6, 12                      \\ 
E.2.7             & None & N/A  \\ 
G.2.3             & SM2.3, SM2.7, CR1.7 & 55, 43, 42              \\ 
E.3.7             & None & N/A                         \\ 
G.1.2             & None & N/A                     \\ 
G.2.6             & None & N/A               \\ 
D.1.3             & CMVM1.1, CMVM1.4, & 92, 73  \\& CMVM2.4 & 34         \\ 
E.3.2             & None & N/A      \\ 
\bottomrule
\end{tabular}
\end{table}

\begin{table}[!htb] 
\caption{BSIMM:  P-SSCRM ranked (lowest to higher) bottom eleven tasks adopted}\label{TAB:BSIMM_Least_Adoption}
\begin{tabular}{l l l}
\toprule
\textbf{P-SSCRM ID} & \textbf{BSIMM15 Mapping} & \textbf{BSIMM15 Adoption \%}                                    \\ \midrule
E.2.6             & None & N/A                 \\ 
P.5.1             & None & N/A                          \\ 
P.3.3            & SE2.4 & 42                      \\ 
G.1.4             & None & N/A \\ 
P.3.5            & None & N/A              \\ 
G.1.3             & None & N/A                         \\ 
D.2.2             & None & N/A                     \\ 
G.1.5             & SE3.6 & 21               \\ 
E.3.9             & None & N/A         \\ 
E.2.3  &   None & N/A     \\ 
P.3.2  &   None & N/A     \\ 
\bottomrule
\end{tabular}
\end{table}

Overall, only 7 of the 30 noteworthy P-SSCRM tasks in Tables \ref{TAB:BSIMM_Top 10Adoption}, \ref{TAB:BSIMM_Least_Adoption}, and \ref{TAB:BSIMM:  Sivana} have a mapping to BSIMM, substantiating a need for a separate effort to analyze trends in the adoption of tasks to reduce software supply chain risk.  As shown in the tables, 5 of these 7 involve a one-to-many mapping between P-SSCRM tasks and BSIMM activities.  The BSIMM15 analysis involves 121 organizations with the P-SSCRM involving only 9.  However, the task/activities that do map do not show a convergence of being among the Top 10 or Bottom 11 between the two frameworks. Overall, the comparison substantiates the need for software supply chain risk management-specific empirical analysis of task adoption.

\begin{table}[tb]
\caption{BSIMM:  P-SSCRM-ranked tasks associated with mitigating current attacker techniques~\cite{hamer2025closing}}\label{TAB:BSIMM:  Sivana}
\begin{tabular}{l l l}
\toprule
\textbf{P-SSCRM ID} & \textbf{BSIMM15 Mapping} & \textbf{BSIMM15 Adoption \%}                                     \\ \midrule
E.3.3              & None & N/A                 \\ 
D.2.1            & None & N/A                          \\ 
E.3.7             & None  & N/A                    \\ 
E.3.6            & None & N/A   \\ 
E.3.11              &  New  & N/A               \\ 
P.5.2             & None & N/A                          \\ 
P.2.1             & AA1.1, AA1.2, & 81, 46\\ & AA2.1, AA3.1 & 31, 17                       \\ 
E.3.4             & None & N/A   \\ 
G.2.6              & None & N/A                       \\ 
D.1.2            & None & N/A   \\ 
\bottomrule
\end{tabular}
\end{table}

\section {Observations} \label{Takeways}

In this section, we share a combination of qualitative and quantitative observations from this study. To start, some of the P-SSCRM tasks, particularly in the Governance Practice and Environment Practice, have been implemented by the organizations in this study for years, well before the current focus on software supply chain security.  The more novel tasks that are focused explicitly on the software supply chain have lower adoption.  

A significant software supply chain attack vector involves attackers leveraging vulnerabilities that are accidentally or intentionally introduced in dependencies, such as components and containers. The adoption of tasks (in Product Practice P3 (0.23), Product Practice P5 (0.24), and Environment Task E.2.2 (0.28)) to select good/secure dependencies, manage their use, and address the vulnerabilities that arise from these dependencies is generally low.  Components and containers may flow freely into an organization without vetting or pre-scanning.  The adoption of tasks that protect an organization from rogue, vulnerable, and unsigned components and containers from entering its development and production environments is emerging.  The average adoption of Tasks P.3.1, P.3.2, P.3.3, and P.3.5 is 0.19, indicating that developers may be able to easily bypass any vetting process if these tasks have been established. In some organizations, unscanned components and containers may be deposited into an organization’s private repository, with the first screening of these new components and containers being done as part of the CI/CD pipeline in a production build.   As organizations compete on delivering functionality, developers feel the need to bring in components and containers without a delay from a vetting process. Tools are needed to automate the vetting process while minimizing the impact on the developer workflow. \textbf{The research indicates that developers can generally bring the dependencies they want to use into the environment and product, bypassing approval and vetted storage processes if these have been established.}  With the growth of the use of the attack vector through the dependencies, organizations are advised to focus on dependency choice and management.

Another major attack vector is attackers utilizing the build infrastructure to launch attacks, such as was done in the SolarWinds and CodeCov attacks. The tasks to close the build infrastructure attack vector (primarily the seven tasks in the Environment practice E2) have an average adoption of 0.42.  \textbf{The research indicates that organizations should continue to mature the use of tasks to prevent more attacks through the build infrastructure.} The average adoption of Tasks E.2.1, E.2.2, and E.2.3 is 0.28, indicating that release policies are not templated or automated, the integrity of build tools, components, and containers is not verified before bringing them into the pipeline, and the build tools are not used to detect vulnerabilities. Some organizations are starting to consider adopting the tasks laid out in the Supply-chain Levels for Software Artifacts (SLSA) \cite{SLSA} framework (which is incorporated in the P-SSCRM). However, adoption is still in the ideation or just emerging stages. Most organizations have not begun to monitor to detect intrusions (D.2.2, 0.18) in the build infrastructure, indicating that an attack through the build infrastructure could go on undetected. None of the organizations has adopted the use of reproducible builds to confirm that no malicious backdoor injections have taken place during the build process.

The interviewees indicated that, once approved, the secure software development life cycle (S-SDLC) and the contract terms process for third-party vendors, as well as the security-related characteristics of open-source components (i.e., whether they are still being maintained), are rarely re-reviewed (P.4.5, 0.58).  For most organizations, the third-party vendor selection process involves completing a survey asking about their S-SDLC practices. Based on the survey answers, vendors may be asked to present a plan to mitigate gaps in their S-SDLC. In general, only vendors of components or tools with high security risk and criticality, such as those that store customer data, undergo an annual re-review. Most interviewees admitted that follow-through on S-SDLC gaps and re-reviews is infrequent. \textbf{The research indicates that, once approved, third-party and open-source system components most often remain in the product.} However, components and containers may be regularly scanned for externally reported vulnerabilities.

The community is having a technical challenge with building a comprehensive asset inventory (G.2.5, 0.41). The goal of maintaining an asset inventory is to facilitate incident response, system analysis, traceability of critical components, and reliable identification when assets need to be modified or decommissioned.  This inventory should include system components, including hardware, software licenses, software versions, direct and transitive component owners, containers, machine names, and network addresses.  Most organizations do not have a systematic approach to maintaining this type of inventory, particularly since many components are hosted in the cloud and/or are ephemeral. \textbf{The research indicates that opportunities exist to provide tool support for systematizing the development of an asset inventory that is dynamic in nature.}

\textbf{The research indicates that organizations are focused on producing an SBOM, but few require SBOMs from their suppliers or use them to respond to security incidents or identify components that need updating or patching.} Only one organization has begun to look at the SBOM of its components; the rest have not started (P.5.1). 

\textbf{The research indicates that the security requirements of third-party suppliers need to be improved.}  The adoption of tasks that require third-party suppliers to employ adequate security measures to protect information, applications, and services provided to the organization is still in progress.  The average adoption of Tasks G3.2, G3.3, G3.4, and G3.5 is 0.58, indicating that one person may make supplier and contractor choices and that contract employees may not be included in session audits. Additionally, suppliers may not need to monitor for or report unauthorized disclosure of information, supply chain threats, or incidents or to declare that a product is heading toward end-of-life.

\textbf{The research indicates that the attack vectors that could lead to unauthorized or accidental access and alteration of project artifacts are still viable.}  The adoption of tasks that make it harder to access project artifacts inappropriately is still in progress.  The average adoption of Tasks E.1.3, E.1.4, E.1.5, and E.1.6 is 0.46. These tasks include the use of multifactor authentication for the source code repository, the use of SSH keys or SSH certificates rather than passwords, the use of branch protection to enforce security policies, and decommissioning end-of-life systems.

\textbf{Attack vectors through the development environments are relatively secure.}   Compared with tasks to protect the attack paths through component/containers and build infrastructure, tasks to protect the software development environment from internal and external threats that can lead to compromise (E.3.1-E.3.8 and E.3.10) are close to being implemented (average adoption of 0.8), with two of the Top 10 Tasks being in this group.  Organizations can continue to secure their development environment through the use of ephemeral credentials (E.3.9, 0.22 adoption).

\begin{tcolorbox}[colback=magenta!5!white,colframe=magenta!75!black]
    \textbf{Key Takeaway:  The most adopted software tasks had been implemented by organizations before the focus on software supply chain security. 
    The tasks that mitigate the novel attack vectors through malicious commits of components and the build infrastructure are in the early stages of adoption.  }
  
\end{tcolorbox}

\section {Limitations} \label{Limitations}

The companies in this study have strong software security management, support a security culture, and have software security technical leadership likely well beyond what is found in a cross-section of organizations in the software industry.  As such, the observations in less focused companies would likely demonstrate lower adoption values.  Through this paper, we establish a baseline for the adoption rates of tasks aimed at reducing software supply chain risk.  We will continue this research with a diverse set of companies internationally, as well as re-interview the original companies to evaluate their adoption longitudinally.  

Interviews provide subjective, self-reported data. The interviews were conducted and scored by one researcher due to NDA constraints. This interviewer has over 35 years of experience with the software industry and in conducting software engineering and software security research studies.  We intentionally interviewed more than one person per company for each task to facilitate cross-checking of answers. Results were reviewed with each company. In the future, we will design NDAs to enable two interviewees and data analysts to enable cross-checking. We will explore how to incorporate automated objective measures, such as those by OpenSSF Scorecard~\cite{Scorecard} metrics, into future work.

Seven of the nine companies have thousands of employees.  Many of our interviews were conducted with centralized services (such as the software security group, vendor management, and risk management). Still, our results for each company may not be generalizable to all teams within the company. 

One person conducted, transcribed, and analyzed the interviews.  The results of the interviews were reviewed with the companies to provide a sanity check on the values assigned to each task.

\section {Ethical Considerations} \label{Ethics}
The interviews were conducted in a business context.  As such, the interview involved signatures on an organization's Non-Disclosure Agreement outlining the voluntary nature of the interviews and the confidentiality of the data. Interviewees were assigned by the main coordinator of the study in each organization.  The motivation for each organization to participate was to obtain a gap analysis of their task adoption compared to the entire set of organizations.

\section{Summary and Future Work}\label{summary}

 The adoption of tasks to protect against and detect malicious infiltration into the software build infrastructure, which could lead to the build and deployment of compromised products, is still in its early stages. 
 Additionally, the adoption of tasks to mitigate the attack vector through components are lagging.  

Our research roadmap includes continuing P-SSCRM-based interviews internationally with additional companies and conducting repeat interviews with previously-interviewed companies.  We are guided in this research roadmap by the success that BSIMM (see Section \ref{BSIMM}) has had and the value BSIMM has provided to organizations and the industry as a whole over the past 17 years. We invite other researchers to join us in a family of research and data sharing around the adoption of the software supply chain security tasks.  We will also work on integrating automated, objective measures of task adoption into our research methodology and will share these methods with the community and collaborators.

\begin{acks}
\label{sec:ack}
This work was supported and funded by the National Science Foundation Grant No. 2207008. Any opinions expressed in this material are those of the author(s) and do not necessarily reflect the views of the National Science Foundation. We thank our colleagues in the Secure Software Supply Chain Center (S3C2) for valuable feedback.
\end{acks}




\bibliographystyle{ACM-Reference-Format}
\bibliography{references}

\end{document}